\documentclass[prl,twocolumn,superscriptaddress,showpacs,amsmath,amssymb]{revtex4-2}
\usepackage{graphicx, color}% Include figure files
\usepackage{dcolumn}% Align table columns on decimal point
\usepackage{bm}% bold math
\usepackage{epstopdf}
\usepackage{physics}
\usepackage[usenames,dvipsnames]{xcolor}
\usepackage{subfigure}
\usepackage{pifont}

\usepackage{verbatim}

\begin{document}
\title{Superconductivity Induced Ferromagnetism In The Presence of Spin-Orbit Coupling} 
\author{Yao Lu}
\email{yao.lu@ehu.eus}
\affiliation{
Centro de F\'{i}sica de Materiales (CFM-MPC), Centro Mixto CSIC-UPV/EHU,  E-20018 San Sebasti\'{a}n, Spain}

\author{I. V. Tokatly}
\email{ilya.tokatly@ehu.es}
\affiliation{Nano-Bio Spectroscopy Group, Departamento de Polímeros y Materiales Avanzados, Universidad del País Vasco, 20018 Donostia-San Sebastián, 
Basque Country, Spain}
\affiliation{IKERBASQUE, Basque Foundation for Science, 48009 Bilbao, Basque Country, Spain}
\affiliation{Donostia International Physics Center (DIPC), 20018 Donostia--San Sebasti\'an, Spain}

\author{F. Sebastian Bergeret}
\email{fs.bergeret@csic.es}
\affiliation{
Centro de F\'{i}sica de Materiales (CFM-MPC), Centro Mixto CSIC-UPV/EHU,  E-20018 San Sebasti\'{a}n, Spain}
 \affiliation{Donostia International Physics Center (DIPC), 20018 Donostia--San Sebasti\'an, Spain}

\date{\today}
\begin{abstract}
We investigate the behavior of magnetic impurities placed on the surface of superconductor thin films with spin-orbit coupling. Our study reveals long-range interactions between the impurities, which decay according to a power law, mediated by the supercurrents. Importantly, these interactions possess a ferromagnetic component when considering the influence of the electromagnetic field, leading to the parallel alignment of the magnetic moments in the case  of two impurities. 
In a Bravais lattice of magnetic impurities, superconductivity facilitates the establishment of ferromagnetic order within specific parameter ranges. These findings challenge the conventional understanding that ferromagnetism and superconductivity are mutually exclusive phenomena. Our theoretical framework provides a plausible explanation for the recently observed remanent flux and transport signature of ferromagnetism in iron-based superconductors, particularly Fe(Se,Te).

\end{abstract}

\maketitle
\paragraph{Introduction.-}Ferromagnetic ordering is often seen as incompatible with conventional superconductivity due to the presence of an effective exchange field in ferromagnets. This exchange field has the effect of breaking up Cooper pairs, which are composed of electrons in a singlet state \cite{anderson1959spin}.
The coexistence of these two orders, however, does exist in hybrid superconductor/ferromagnet (S/F)  structures\cite{buzdin2005proximity,bergeret2005odd}.
In these structures, the proximity effect plays a crucial role in enabling the coexistence of superconductivity and ferromagnetism.
 Singlet pairs can be transformed into triplet pairs through the exchange field of the F region. As a result, a local magnetic moment is produced, extending over distances of the order of superconducting coherence length, $\xi_s$. This phenomenon is referred to as the magnetic or inverse proximity effect\cite{bergeret2004induced,xia2009inverse}. 
 The generated magnetic moment is oriented in the opposite direction to the  magnetization of the ferromagnetic region. In the case of a small ferromagnetic island, this results in the screening of its magnetic moment\cite{bergeret2004spin}.
If a second ferromagnetic region (F region) is positioned at a distance smaller than $\xi_s$ from the first ferromagnet, the energetically favorable arrangement is an anti-parallel orientation of the magnetizations of the two F regions. 
This anti-parallel alignment serves as the basis for the FSF superconducting spin-valve\cite{de1966coupling,tagirov1999low,izyumov2002competition,ojajarvi2022dynamics}.
 The studies on FSF structures with conventional superconductors indicate an anti-ferromagnetic coupling between the magnets, mediated by the inverse proximity effect. This coupling strength decreases exponentially with the distance between the ferromagnetic regions.
\begin{figure}[t]
\centering
\includegraphics[width = 0.7\columnwidth]{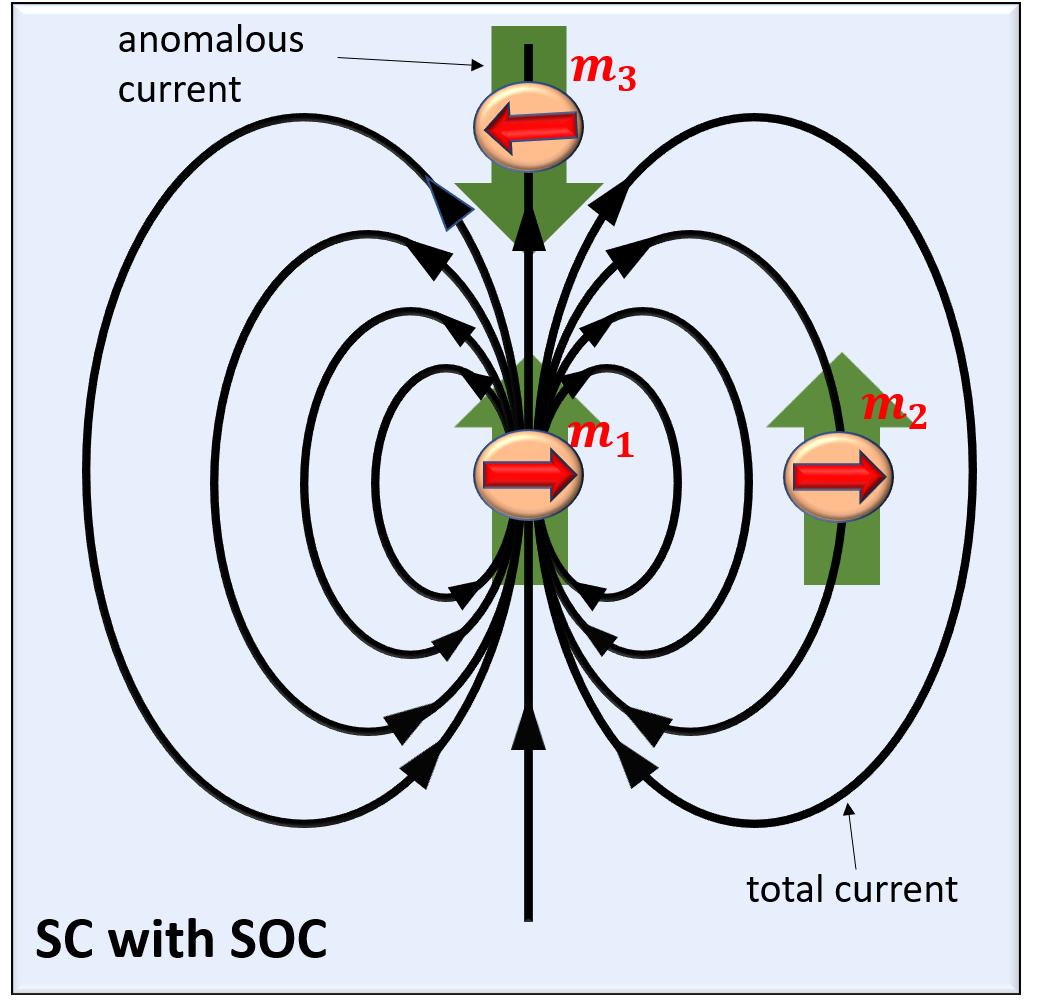}
\caption{Schematic picture of magnetic impurities on top of a 
superconductor thin film with spin-orbit coupling. The red arrows represent the magnetization of the impurities. The green arrows are the exchange field-induced localized anomalous currents. The black loop represents the total current induced by the exchange field, phase gradient, and electromagnetic field. 
%The magnetic impurities interact with each other through the supercurrent.
}
\label{Fig:Schematic}
\end{figure}
 This situation changes   in thin superconducting films  with  spin-orbit coupling (SOC). 
 The combination of the exchange field generated by a magnetic impurity,  $m_1$ in Fig. ~\ref{Fig:Schematic}) , and the SOC   results in the spontaneous generation of anomalous currents through the spin-galvanic effect. In the case of a Rsshba SOC  they  flow 
 perpendicular to the magnetization (green arrows in Fig.~\ref{Fig:Schematic}) 
 \cite{edelstein1995magnetoelectric,konschelle2015theory}.
 These anomalous currents are spatially localized\cite{bergeret2020theory} over the coherence length from the impurity.  Charge conservation 
implies the emergence of a phase gradient , ensuring $\boldsymbol{\nabla}\cdot\boldsymbol{j}=0$ and the appearance of circulating currents \cite{pershoguba2015currents} (black arrows in Fig.~\ref{Fig:Schematic})
 If we assume that a  magnetic  moment 
$\boldsymbol{m}_1$ points  in the positive $x$ direction (Fig.~\ref{Fig:Schematic}),  then it  generates a non-local circular supercurrent which flows in the negative $y$ direction at the position of the  magnetic impurity $\boldsymbol{m}_2$  and to the positive $y$ direction at the position of  $\boldsymbol{m}_3$.
The orientations of  $\boldsymbol{m}_{2,3}$
 are determined by minimization of the free energy: to reduce the kinetic energy of the superflow they will generate anomalous currents that suppress the currents  induced by $\boldsymbol{m}_1$. Consequently, $\boldsymbol{m}_2$  will point in the  positive $x$ direction and $\boldsymbol{m}_3$
to the negative $x$ direction. In other words,  the supercurrent mediated magnetic interaction is ferromagnetic between  $\boldsymbol{m}_1$ and $\boldsymbol{m}_2$ while it is antiferromagnetic for $\boldsymbol{m_1}$ and $\boldsymbol{m}_3$. Thus,  in general, for two magnetic impurities,   $\boldsymbol{m}_1$ and $\boldsymbol{m}_2$ the  magnetic interaction will have the form: 
\begin{equation}
    \label{eq:generalF}
F_I=J_{\perp}\boldsymbol{m}_{1\perp}\boldsymbol{m}_{2\perp}-J_{\parallel}\boldsymbol{m}_{1\parallel}\boldsymbol{m}_{2\parallel}\;, 
\end{equation}
where $\perp$ ($\parallel$) denotes the component in the direction perpendicular (parallel) to $\boldsymbol{r}=\boldsymbol{r}_1-\boldsymbol{r}_2$,  and both $J{\perp}$ and $J_{\parallel}$ are positive.
Previous studies  on Rashba superconductors\cite{pershoguba2015currents, mal2018nonexponential}, specifically regarding the current distribution around a magnetic impurity and the induced magnetic interaction  have obtained an interaction resembling the 2D dipole-dipole interaction (DDI)  form with $J_{\perp}=J_{\parallel}$, which does not result in either a ferromagnetic or an anti-ferromagnetic ground state for two impurities. However, those studies
have  neglected the influence of the electromagnetic (EM) field. On the other hand, we know from Pearl's   seminal work \cite{pearl1964current}  that the EM field plays a crucial role in determining the current distribution in conventional superconducting thin films.
This leads to the natural question of the effect of the EM field on the magnetic coupling between impurities in a superconductor with spin-orbit coupling (SOC).

In this work, we present a theory
elucidating the impact of the electromagnetic field on the magnetic coupling between impurities in a superconductor with SOC.
We demonstrate that the presence of EM fields alters drastically the spatial dependence of the couplings $J_{\perp(\parallel)}$. We establish that the supercurrent-mediated magnetic interaction exhibits the form of a DDI that is generated by the so-called Keldysh potential \cite{Keldysh1979,cudazzo2011dielectric}, and interpolates between the 2D and 3D DDI. It can also be viewed as a 2D DDI combined with a ferromagnetic interaction, leading to a ferromagnetic ground state for two impurities. Furthermore, we emphasize the crucial role of the electromagnetic field in a 2D impurity lattice. Without the electromagnetic field, the interaction energy density becomes unphysically divergent as the system size approaches infinity. However, when the electromagnetic field is taken into account, the energy density converges in the limit of large system size. 
%ensuring its  extensivity.
In the remainder of this paper, we provide a detailed  derivation of these results and discuss recent experimental results suggesting a superconducting-induced ferromagnetic order of impurities in  Fe(SeTe) \cite{xiang2023observation}. 
\paragraph{Theory.-} 
We first consider magnetic impurities on top  of a two-dimensional superconducting system with SOC, which can, for example, be a thin film on a substrate or superconductivity induced at the surface of a topological insulator. We assume all the magnetic impurities are polarized in the in-plane directions. In these systems, the inversion symmetry is broken because of the presence of the intrinsic polar vector -- the normal $\hat z$ to the transport plane. Additionally, the exchange field $\boldsymbol{h}$ induced by the magnetic impurity locally breaks the time-reversal symmetry. The breaking of these two symmetries allows for the existence of a spontaneous current, known as the anomalous current, given  by\cite{bergeret2015theory,strambini2020josephson} $\boldsymbol{j}=-e^2D\boldsymbol{a}$, where $D$ is the 2D superfluid weight and $\boldsymbol{a}$ is the effective gauge field $\boldsymbol{a}=\frac{1}{e}\Gamma \boldsymbol{h}\times\hat{z}$. Here  $\Gamma$ is a constant that depends on the details of the system. For example, in a superconductor with strong Rashba SOC $\Gamma=\alpha/v_F^2$ (see section 4 of supplementary material \cite{supplementary}), where $\alpha$ is the SOC strength; or  $\Gamma=1/v_F$ in a Dirac material\cite{zyuzin2016josephson,kokkeler2022field}.

The total free energy change due to the supercurrent is given by 
\begin{equation}
    F=\int d^3\boldsymbol{r}\frac{1}{8}D\left[\boldsymbol{\nabla}\phi-2e\boldsymbol{A}-2e\boldsymbol{a}\right]^2\delta(z)+\frac{1}{2\mu_0}\boldsymbol{B}^2\; . \label{eq:energy1}
\end{equation}
The first term on the right-hand side is the free energy of the superconductor $F_{SC}$ and the second term is the electromagnetic field contribution $F_{EM}$,
where $\mu_0$ is the magnetic constant, $e$ is the electron charge, $\boldsymbol{A}$ is the vector potential of the electromagnetic field $\boldsymbol{B}=\boldsymbol{\nabla}\times\boldsymbol{A}$.   We assume the superconductor film is located at $z=0$.  Note that $\boldsymbol{A}$ lives in 3 dimensions while $\boldsymbol{\nabla}\phi$ and $\boldsymbol{a}$ are defined in the 2D superconductor.  Taking the derivative of $F_{SC}$ with respect to $-\boldsymbol{A}$, one obtains the 2D supercurrent flowing in the plane of the superconductor, 
\begin{equation}
    \boldsymbol{j}=\frac{1}{2}eD\left[\boldsymbol{\nabla}\phi-2e\boldsymbol{A}-2e\boldsymbol{a}\right]\delta(z).\label{eq:current0}
\end{equation}
Minimizing the free energy with respect to $\phi$ gives the charge conservation $\boldsymbol{\nabla}\cdot\boldsymbol{j}=0$. In the following, we choose the Coulomb gauge $\boldsymbol{ \nabla}\cdot\boldsymbol{A}=0$. This implies that the phase gradient cancels the longitudinal part $\boldsymbol{a}_l$  of the effective gauge field, which  is written in the momentum $\boldsymbol{q}$ space as,  $\boldsymbol{q}\phi=2e\boldsymbol{a}_l=2e\boldsymbol{q}(\boldsymbol{q}\cdot\boldsymbol{a})/q^2$. The current of Eq.~\eqref{eq:current0} is then fully determined by the trasverce component $\boldsymbol{a}_t=\boldsymbol{a}-\boldsymbol{a}_l$, that is, $\boldsymbol{j}=-e^2D(\boldsymbol{A} + \boldsymbol{a}_t)$.

By minimizing the total free energy with respect to the vector potential $\frac{\partial F}{\partial\boldsymbol{A}}=0$, we obtain the Maxwell equation $\mu_0\boldsymbol{j}=\boldsymbol{\nabla}\times\boldsymbol{B}$, which in the Coulomb gauge reads,
\begin{equation}\label{Maxwell}
\boldsymbol{\nabla}^2\boldsymbol{A}=e^2\mu_0D\left[\boldsymbol{A} + \boldsymbol{a}_t\right]\delta(z).
\end{equation}
The solution of this equation for a given distribution of the effective gauge field $\boldsymbol{a}(\boldsymbol{r})$ determines the induced vector potential and the charge current. 
At the solution point, by substituting the Maxwell equation back into Eq.~\eqref{eq:energy1}, we get the change of free energy due to supercurrents generated by the external exchange field, (see section 1 of the supplementary material \cite{supplementary})
\begin{equation}\label{eq:simplifiedenergy}
    F=-\frac{1}{2}\int d^2\boldsymbol{r}\,\boldsymbol{j}\boldsymbol{a}. 
\end{equation}
Remarkably, the  free energy  is determined by the supercurrent in the regions in which  $\boldsymbol{a}(\boldsymbol{r})$ is finite, {\it i.e.} the regions where the exchange field $\boldsymbol{h}(\boldsymbol{r})$ is finite. 
\begin{figure}[t]
\centering
\includegraphics[width = \columnwidth]{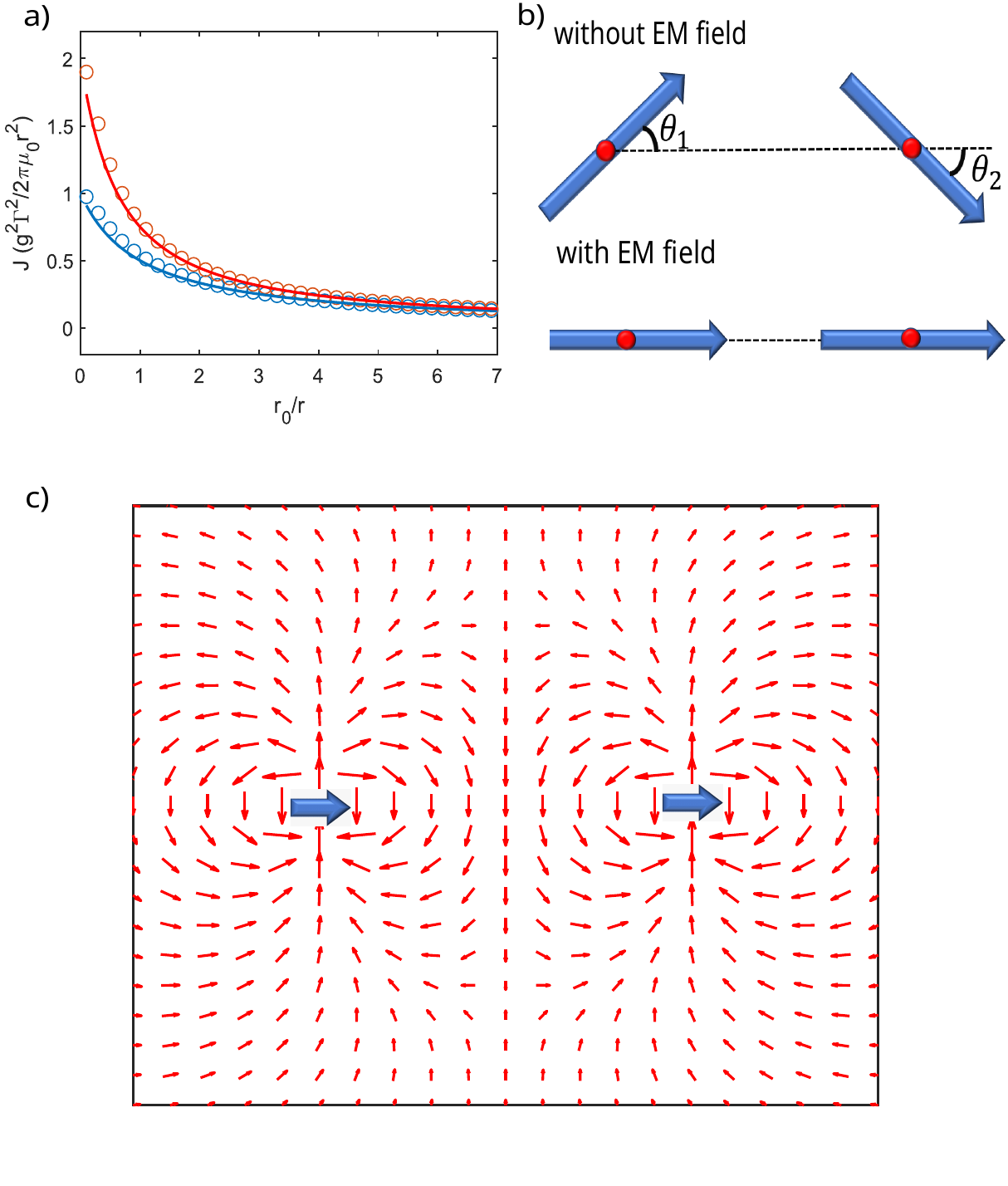}
\caption{a) Superconductivity-induced magnetic interaction as a function of Pearl length. The red and blue colors denote $J_{\parallel}$ and $J_{\perp}$, respectively. The circles are the exact results calculated from Eq. (\ref{eq:exact}) and the lines are the approximate values obtained from Eq. (\ref{eq:approximate}). b) Ground states of two magnetic impurities with and without the EM field. c) Distribution of the supercurrent induced by the two magnetic impurities.   The distance between the two impurities is $r=r_0/2$.}\label{Fig:2FI}
\end{figure}

We now consider a set of magnetic regions (impurities) located in the superconducting plane at the points $\boldsymbol{r}_i$, such that the distance between the regions is much larger than their size. In this case, the distribution of the induced supercurrents almost everywhere as well as the change of the free energy become independent on the size/shape of the impurities and the corresponding exchange field can be approximated as $\boldsymbol{h}(\boldsymbol{r})=\sum_i J_0\boldsymbol{m}_i\delta(\boldsymbol{r}-\boldsymbol{r}_i)$. Here $\boldsymbol{m}_i$ is the total magnetization (spin) of the ith impurity, and $J_0$ is the electron-impurity exchange coupling.

By solving the Maxwell-London equation (\ref{Maxwell}) and inserting the supercurrent $\boldsymbol{j}=-e^2D(\boldsymbol{A} + \boldsymbol{a}_t)$ into Eq.~\eqref{eq:simplifiedenergy} we can identify the part of the free energy responsible for the supercurrent-induced long-range magnetic interaction
(see section 2 of supplementary material \cite{supplementary}):
\begin{equation}\label{interaction}
    F_I=-\frac{Z}{2}\sum_{i\ne i}(\boldsymbol{m}_i\cdot\nabla)(\boldsymbol{m}_j\cdot\nabla)V(r_{ij}) = \frac{1}{2}\sum_{i\ne j}F_I^{ij},
\end{equation}
where $Z=DJ_0^2\Gamma^2/2\pi$, $r_{ij}=|\boldsymbol{r}_i-\boldsymbol{r}_j|$ is the distance between magnetic impurities, and  $V(r)$ is the dimensionless Keldysh potential,
\begin{equation}\label{Keldysh}
V(r)=\frac{\pi}{2}\left[H_0\left(\frac{r}{r_0}\right)-Y_0\left(\frac{r}{r_0}\right)\right].
\end{equation}
Here $H_0$ is the Struve function, $Y_0$ is the second kind Bessel function, and $r_0$ is the Pearl length $r_0=2/e^2D\mu_0$ \cite{pearl1964current}. 

The appearance of the Keldysh potential is quite remarkable. Usually, it describes the electrostatic potential of a point charge confined to a polarizable insulating plane \cite{Keldysh1979,cudazzo2011dielectric}. It interpolates between the 2D ($V\sim -\ln r$) and the 3D ($V\sim {1}/{r}$) forms of the Coulomb potential with the crossover scale given by $r_0$ in Eq.~\eqref{Keldysh}. Here it plays a similar role by interpolating between the 2D and 3D form of the induced DDI between the impurity spins.

Working out the derivatives in the pairwise part $F_I^{ij}$ of the interaction energy Eq.~\eqref{interaction} we find,

\begin{equation}
    \label{eq:generalF2}
F_I^{ij}=J_{\perp}(r_{ij})\boldsymbol{m}_{i\perp}\boldsymbol{m}_{j\perp}-J_{\parallel}(r_{ij})\boldsymbol{m}_{i\parallel}\boldsymbol{m}_{j\parallel}\; , 
\end{equation}

with 
\begin{equation}\label{eq:exact}
    J_{\perp}(r)=-Z\frac{1}{r}\frac{dV}{dr},\quad J_{\parallel}(r)=Z\frac{d^2V}{dr^2}.
\end{equation}
To gain insight into  the $r$-dependence, it is instructive to use a highly accurate representation of
 $V(r)$ in terms of elementary functions \cite{cudazzo2011dielectric}, which yields
\begin{equation} \label{eq:approximate}
    J_{\perp}=Z\frac{r_0}{r^2(r+r_0)}, \quad J_{\parallel}=Z\frac{r_0(2r+r_0)}{r^2(r+r_0)^2}.
\end{equation}
The interaction as a function of the  Pearl length is shown in Fig.~\ref{Fig:2FI}a.  The important general property is that for any finite  $r_0$, one obtains $J_{\parallel}>J_{\perp}$.

Let us analyze the case of two impurities. For convenience, we write the interaction as 
\begin{equation}
    F_I=\frac{J_{\perp}+J_{\parallel}}{2}\left(m_{1\perp}m_{2\perp}-m_{1\parallel}m_{2\parallel}\right)-\frac{J_{\parallel}-J_{\perp}}{2}\boldsymbol{m}_1\cdot\boldsymbol{m}_2.
\end{equation}
The first term on the right-hand side alone does  not generate a difference in the free energies of the ferromagnetic and anti-ferromagnetic states but rather leads to degenerate ground states  with $\theta_1=-\theta_2$, where $\theta_1$ and $\theta_2$ are the angles between the impurities magnetic moments  and $\boldsymbol{r}$ as shown in Fig.~\ref{Fig:2FI}b. The second term, with the  form of isotropic ferromagnetic interaction, breaks the ground state degeneracy, resulting in a ferromagnetic ground state. This ferromagnetic interaction stems from the $\boldsymbol{B}^2$ term in the free energy, which was ignored in  previous works \cite{pershoguba2015currents,mal2018nonexponential}.  In superconductors with large Rashba SOC $\Gamma$ and $D$ are given by $\Gamma=\alpha/v_F^2$, and $D=\frac{1}{2}\pi v_F^2N_0T\sum_n\frac{\Delta^2}{(\omega_n^2+\Delta^2)^{3/2}}$ (see section 3 of supplementary material \cite{supplementary}). In the limit where the EM field can be neglected $\mu_0\rightarrow 0$ and $r_0\rightarrow \infty$, we get $J_{\parallel}=J_{\perp}$ and recover the result of Ref.~\cite{mal2018nonexponential}.

The supercurrent $\boldsymbol{j}(\boldsymbol{r})$ induced by a magnetic impurity with magentization $\boldsymbol{m}$ at the origin, takes the form \cite{supplementary}

\begin{equation}
    \boldsymbol{j}(\boldsymbol{m},\boldsymbol{r})=\frac{1}{\Gamma  
 J_0}\left[J_{\parallel}\boldsymbol{m}\times\hat{z}-(J_{\parallel}+J_{\perp})(\boldsymbol{m}\times\hat{z}\cdot\hat{r})\hat{r}\right],
\end{equation}
where $\hat{r}\equiv\boldsymbol{r}/|\boldsymbol{r}|$. Due to the linearity of the problem, the total current induced by all impurities is given by sum $\boldsymbol{j}_{\rm tot}=\sum_i\boldsymbol{j}(\boldsymbol{m}_i,\boldsymbol{r}-\boldsymbol{r}_i)$. The current distribution for two impurities is shown in Fig.~\ref{Fig:2FI}c.  Experimentally, the supercurrent distribution can be determined  by measuring the local  current-induced magnetic field\cite{lin2023direct}
%using scanning
%superconducting quantum interference device (SQUID) microscopy. 
The magnetic field $\boldsymbol{B}=\boldsymbol{\nabla}\times\boldsymbol{A}$ induced by one impurity at the origin, and evaluated at $z=0$ reads,
\begin{equation}
    \boldsymbol{B}(\boldsymbol{m},\boldsymbol{r})=\frac{\Gamma J_0}{4\pi}\left[\frac{2}{(r+r_0)^3}+\frac{2r+r_0}{r^2(r+r_0)^2}\right](\boldsymbol{m}\cdot\hat{r})\hat{z}.
\end{equation}
The total magnetic field at $z=0$ is then $\boldsymbol{B}_{\rm tot}=\sum_i\boldsymbol{B}(\boldsymbol{m}_i,\boldsymbol{r}-\boldsymbol{r}_i)$. %Thus one can probe the state of two magnetic impurities using SQUID \cite{lin2023direct}.

\paragraph{Superconductivity induced ferromagnetism in a 2D Bravais lattice.-}
\begin{figure}[t]
\centering
\includegraphics[width = 1\columnwidth]{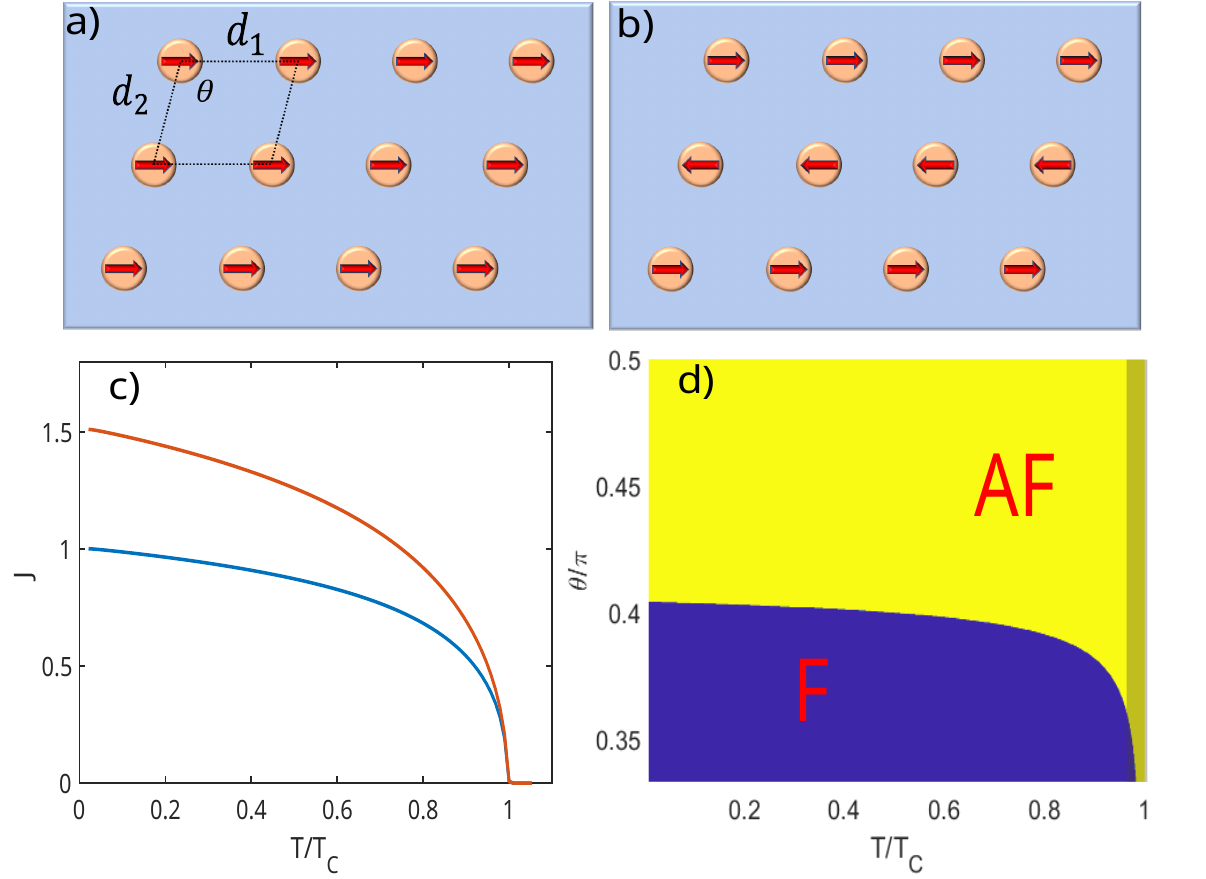}
\caption{a-b) Schematic picture of the ferromagnetic and the layered anti-ferromagnetic states of a Bravais magnetic impurity lattice.  c) Temperature dependence of the interaction strength. The red line and blue line denote $J_{\parallel}$ and $J_{\perp}$, respectively in units of  $J_{\perp}(T=0)$. The distance between the two impurities is $r=r_0(T=0)/2$. d) Phase diagram for a 2D impurity lattice. The lattice constant here is $d_1=d_2=r_0(T=0)/50$. When $T$ is close to $T_c$ (grey shaded region), the Pearl length becomes  too long and there might be numerical uncertainties due to the finite size effect. }\label{Fig:2Dlattice}
\end{figure}
Next, we consider an infinite Bravais lattice of magnetic impurities, as shown in Fig.~\ref{Fig:2Dlattice}a. Below we assume $d_1=d_2$ and the shape of the lattice is controlled by the angle $\theta$. It is a triangular lattice when $\theta=\pi/3$ and a square lattice when $\theta=\pi/2$. Here we concentrate on the case where $\theta$ is between $\pi/3$ and $\pi/2$. In this system, the interaction energy density is given by $E=\frac{1}{2V}\sum_{i\neq j}F_{I}^{ij}$, where $V$ is the area of the 2D lattice.
We notice that the inclusion of the EM field is crucial for computing energy. Otherwise, the interaction $F_I^{ij}$ scales as $1/r^2$ and the energy density unphysically diverges in the thermodynamics limit.
With the screening effect of the EM field, according to Eq. (\ref{eq:approximate}), $F_I^{ij}$ crosses over from the 2D to 3D DDI and decays as $1/r^3$ at $r\rightarrow \infty$, resulting in convergent and extensive energy.

The 3D DDI induced ordered state in a 2D electric dipole lattice has been studied in \cite{rozenbaum19910rientati}. It has been shown that a ferroelectric state forms in a triangular lattice, while a square lattice favors a layered anti-ferroelectric state. Since the superconductivity-induced magnetic interaction has a form between 2D and 3D DDI, we expect similar ground states in our model. By minimizing the energy  we find that at $T=0$ the ground state can be either a ferromagnetic state,   Fig.~\ref{Fig:2Dlattice}a,  induced by the net ferromagnetic interaction $\frac{J_{\parallel}-J_{\perp}}{2}$, or layered anti-ferromagnetic state, Fig.~\ref{Fig:2Dlattice}b,  due to the 2D DDI $\frac{J_{\parallel}+J_{\perp}}{2}$.  At finite temperatures, the order parameter $\Delta(T)$ needs to be determined  self-consistently.
The superfluid weight $D$ and the Pearl length $r_0$  are calculated using  $\Delta(T)$. The corresponding temperature dependence of $J_{\parallel}$ and $J_{\perp}$ is shown in Fig.~\ref{Fig:2Dlattice}c.  
With increasing temperature, the superfluid weight is decreased, leading to the suppression of the magnetic interaction. In addition, the relative value of the net ferromagnetic interaction $\frac{J_{\parallel}-J_{\perp}}{2}$ compared with the 2D DDI $\frac{J_{\parallel}+J_{\perp}}{2}$ becomes smaller at higher temperatures, suggesting a transition to an antiferromagnetic phase at some finite temperature.  The phase diagram obtained numerically  is shown in Fig.~\ref{Fig:2Dlattice}d. At $T=0$, the triangular lattice ($\theta=\pi/3$) has a ferromagnetic ground state, while the square lattice is antiferromagnetic.  By setting $\mu_0\rightarrow 0$, we find that the ground state is always an antiferromagnetic state without the magnetic field.

\paragraph{Conclusion.-}To conclude, we have demonstrated that superconductivity  in materials with SOC can induce long-range magnetic interaction. The effects of London-Pearl screening are crucial for proving this result. The exact analytic solution shows that the induced
% In order to prove this result it is crucial to take into account the EM field. By analytically solving the exact current distribution in the superconductor
% we find that this 
magnetic interaction has a form of 2D DDI combined with a ferromagnetic coupling.  We  also demonstrate that in an oblique Bravais lattice of magnetic impurities, the superconductivity can induce a ferromagnetic state in certain parameter regimes. To the best of our knowledge, these results provide the first example of singlet superconductivity-induced ferromagnetic coupling.

In principle, our predictions  can be tested in any superconductor that exhibits a sizable Rashba SOC or in a Dirac superconductor  with magnetic impurities. %One approach to confirm these predictions is by measuring the local magnetic moment\YL{Now we cite more papers, the magnetization is measured in several different methods, not only measuring the local magnetic moment}. 
Notably, recent experiments \cite{xiang2023observation,qiu2023concurrent,zaki2021time,mclaughlin2021strong,farhang2023revealing}, provides compelling evidence of observing such a ferromagnetic state. Specifically, in Ref.~\cite{xiang2023observation}, a hysteretic magnetization was observed in Fe(SeTe) with Fe impurities. The magnetization disappears at temperatures above the superconducting critical temperature, indicating that the ferromagnetism is induced by the superconductivity. This observation can be considered as evidence of supercurrent-mediated magnetic interaction in the presence of surface Dirac states.
The realistic values for the parameters of the surface Dirac band in Fe(Te,Se)  \cite{xiang2023observation} are given by \cite{jiang2019quantum,thampy2012friedel}: $v_F=0.216eV\AA$, $E_F=4.5meV$, $\Delta=1.5meV$ and ${\cal J}_0=50meV$, $|\boldsymbol{m}|=5$ and the distance between the nearest impurities is $d=2nm$. Here ${\cal J}_0$ is the exchange interaction defined in a lattice model. In our continuous model, the exchange interaction is $J_0={\cal J}_0{\cal A}$, where ${\cal A}$ is the effective area  of the impurity which is assumed to be the lattice constant square ${\cal A}=(0.4nm)^2$. We obtain at $T=0$ the supercurrent-mediated magnetic interaction  is roughly $2meV$, of the same order as the superconducting gap. Thus, the thermal fluctuation can be neglected for low enough temperatures $T\ll \Delta$.  Note that the effective magnetic  interaction is proportional to ${\cal A}^2$, so it can be easily  enhanced by increasing the size of the magnetic  impurity, for example using  islands of a ferromagnetic insulator with a large size.  According to Ref.~\cite{qiu2023concurrent,farhang2023revealing}, the ferromagnetism dwells on the surface of the superconductor, supporting our theory that the SOC is crucial for the formation of the ferromagnetic state.

\begin{acknowledgments}
      {\bf Acknowledgements}
Y.L. and F.S.B. acknowledge  financial support from Spanish AEI through project PID2020-114252GB-I00 (SPIRIT), TED2021-130292B-C42, and  the Basque Government through grant IT-1591-22 and IKUR strategy program.  
I.V.T. acknowledges support by
Grupos Consolidados UPV/EHU del Gobierno Vasco
(Grant IT1453-22) and by the grant PID2020-112811GB-I00 funded by MCIN/AEI/10.13039/501100011033.
\end{acknowledgments}

\onecolumngrid
\pagebreak
\clearpage

\setcounter{equation}{0}
\setcounter{figure}{0}
\setcounter{table}{0}
\setcounter{page}{1}
\renewcommand{\thefigure}{S\arabic{figure}}
\renewcommand{\theequation}{S\arabic{equation}}

\renewcommand{\theequation}{S\arabic{equation}}
\renewcommand{\thesection}{S\arabic{section}}
\renewcommand{\thefigure}{S\arabic{figure}}

\begin{center}
	\textbf{\large Supplemental material for Superconductivity Induced Ferromagnetism In The Presence of Spin-Orbit Coupling}
\end{center}
In this supplementary material, we present the derivations of the simplified free energy,  the spatial distribution of supercurrents induced by magnetic impurities, the supercurrent-mediated magnetic interaction, the superfluid weight of a 2D Rashba superconductor and the $\Gamma$ parameter for a 2D Rashba superconductor. 
\section{S1. Derivation of the free energy}
We start with the full expression of the free energy, which is Eq. (1) in the main text

\begin{equation}
    F=\int d\boldsymbol{r}\left[\frac{1}{8}D\left(\boldsymbol{\nabla}\phi-2e\boldsymbol{A}-2e\boldsymbol{a}\right)^2\delta(z)+\frac{1}{2\mu_0}\boldsymbol{B}^2\right], \label{eq:energy1}
\end{equation}
where $\mu_0$ is the magnetic constant, $e$ is the electron charge, $\boldsymbol{A}$ is the vector potential of the electromagnetic field and $\boldsymbol{B}=\boldsymbol{\nabla}\times\boldsymbol{A}$ is the strength of the magnetic field.  $z$ is the out-of-plane direction and we assume the superconductor film is located at $z=0$.  $D$ is the 2D superfluid weight. $\boldsymbol{a}$ is the effective gause field induced by the magnetic impurities

\begin{equation}
    \boldsymbol{a}=\Gamma J_0\boldsymbol{m},\label{eq:effective gauge field}
\end{equation}
where $\Gamma$ is a constant that depends on the spin-orbit coupling. $J_0$ is the exchange interaction and $\boldsymbol{m}$ is the magnetization of the impurities. The first term on the right-hand side of Eq. (\ref{eq:energy1}) is the free energy of the superconductor $F_{SC}$ and the second term in the electromagnetic field contribution $F_{EM}$.  Taking the derivative of $F_{SC}$ with respect to $\boldsymbol{A}$, one obtain the supercurrent density

\begin{equation}
    \boldsymbol{j}=-\frac{\partial F_{SC}}{\partial A}=\frac{1}{2}eD\left[\boldsymbol{\nabla}\phi-2e\boldsymbol{A}-2e\boldsymbol{a}\right]\delta(z).\label{eq:current0}
\end{equation}

The $F_{EM}$ term can also be written in terms of the vector potential $\boldsymbol{A}$ 

\begin{eqnarray}
  \frac{1}{2}\int d\boldsymbol{r}  \boldsymbol{B}^2&=&\frac{1}{2}\int d\boldsymbol{r}\left(\boldsymbol{\nabla}\times\boldsymbol{A}\right)\cdot\boldsymbol{B}\nonumber\\  &=&\frac{1}{2}\int d\boldsymbol{r}\boldsymbol{A}\cdot\left(\boldsymbol{\nabla}\times\boldsymbol{B}\right).\label{eq:B square}
\end{eqnarray}

In the second line of Eq. (\ref{eq:B square}), we have used integration by part and the fact that $\boldsymbol{A}(r\rightarrow\infty)\rightarrow 0$. Substituting Eq. (\ref{eq:B square}) into Eq. (\ref{eq:energy1}), we have

\begin{equation}
     F=\int d\boldsymbol{r}\left[\frac{1}{8}D\left(\boldsymbol{\nabla}\phi-2e\boldsymbol{A}-2e\boldsymbol{a}\right)^2\delta(z)+\frac{1}{2\mu_0}\boldsymbol{A}\cdot\left(\boldsymbol{\nabla}\times\boldsymbol{B}\right)\right]. \label{eq:energy2}
\end{equation}

Minimization of the free energy repect to $\boldsymbol{A}$,  $\frac{\partial F}{\partial\boldsymbol{A}}=0$, gives the Maxwell equation 
\begin{equation}
 \boldsymbol{j}=\frac{1}{\mu_0}\boldsymbol{\nabla}\times\boldsymbol{B}. \label{eq:maxwell}
\end{equation}
 Substituting Eq. (\ref{eq:maxwell}) and Eq. (\ref{eq:current0}) into Eq. (\ref{eq:energy2}), we have

\begin{eqnarray}
    F&=&\frac{1}{2}\int d\boldsymbol{r}\left[\frac{1}{2e}\boldsymbol{j}\left(\boldsymbol{\nabla}\phi-2e\boldsymbol{A}-2e\boldsymbol{a}\right)+\boldsymbol{A}\cdot\boldsymbol{j}\right]\nonumber\\
    &=&\frac{1}{2}\int d\boldsymbol{r}\frac{1}{2e}\boldsymbol{j}\left(\boldsymbol{\nabla}\phi-2e\boldsymbol{a}\right). \label{eq:energy3}
\end{eqnarray}
The first term on the right-hand side vanishes 

\begin{eqnarray}
    F_1&=&\frac{1}{4e}\int d\boldsymbol{r}\,\boldsymbol{j}\boldsymbol{\nabla}\phi \nonumber\\
       &=&\frac{1}{4e}\int d\boldsymbol{r}\,\left[\boldsymbol{\nabla}\left(\boldsymbol{j}\phi\right)-\phi\boldsymbol{\nabla}\phi\right]\nonumber\\
       &=&0.
\end{eqnarray}
Here we have used $\boldsymbol{\nabla}\boldsymbol{j}=0$ and $\boldsymbol{j}(r\rightarrow 0)\rightarrow 0$. Thus the free energy is given by

\begin{equation}
    F=-\frac{1}{2}\int d\boldsymbol{r}\,\boldsymbol{j}\boldsymbol{a}.\label{eq:energyf}
\end{equation}
This is equation (4) in the main text.

\section{S2. Derivation of the supercurrent and the  interaction energy}
Here we consider two magnetic impurities $\boldsymbol{m}_1$ and $\boldsymbol{m}_2$ located at $\boldsymbol{r}_1$ and $\boldsymbol{r}_2$ respectively. The free energy can be written as

\begin{equation}
    F=-\frac{1}{2}\int d\boldsymbol{r} \left(\boldsymbol{j}_1+\boldsymbol{j}_2\right)\left[\boldsymbol{a}_1\delta(\boldsymbol{r}-\boldsymbol{r}_1)+\boldsymbol{a}_2\delta(\boldsymbol{r}-\boldsymbol{r}_2)\right],
\end{equation}
where $\boldsymbol{a}_1$ and $\boldsymbol{a}_2$ are the effective gauge fields induced by $\boldsymbol{m}_1$ and $\boldsymbol{m}_2$, respectively $\boldsymbol{a}_{1,2}=\Gamma J\boldsymbol{m}_{1,2}$. $\boldsymbol{j}_1$ ($\boldsymbol{j}_2$) is the supercurrent induced by $\boldsymbol{m}_1$ ($\boldsymbol{m}_2$). Working out the integral we obtain

\begin{equation}
    F=-\frac{1}{2}\left[\boldsymbol{j}_1(\boldsymbol{r}_2)\boldsymbol{a}_2+\boldsymbol{j}_2(\boldsymbol{r}_1)\boldsymbol{a}_1+j_1(r_1)\boldsymbol{a}_1+\boldsymbol{j}_2(\boldsymbol{r}_2)\boldsymbol{a}_2\right].\label{eq:energy3}
\end{equation}
The first two terms in Eq. (\ref{eq:energy3}) represent the interaction energy

\begin{equation}
    F_I=-\frac{1}{2}\left[\boldsymbol{j}_1(\boldsymbol{r}_2)\boldsymbol{a}_2+\boldsymbol{j}_2(\boldsymbol{r}_1)\boldsymbol{a}_1\right]=\boldsymbol{j}_1(\boldsymbol{r}_2)\boldsymbol{a}_2.\label{eq:interactionenergy}
\end{equation}
Here we have used $\boldsymbol{j}_1(\boldsymbol{r}_2)\boldsymbol{a}_2=\boldsymbol{j}_2(\boldsymbol{r}_1)\boldsymbol{a}_1$. The main task is to solve the current distribution $\boldsymbol{j}_1$.  Without loss of generality, we place $\boldsymbol{m}_1$ at the origin $\boldsymbol{r}_1=0$ and calculate the current at position $\boldsymbol{r}$. We begin with the expression of the current density

\begin{equation}
    \boldsymbol{j}_1=\frac{1}{2}eD\left[\boldsymbol{\nabla}\phi-2e\boldsymbol{A}-2e\boldsymbol{a}_1\right]\delta(z).\label{eq:current2}
\end{equation}
Here the effective gauge field $\boldsymbol{a}_1$ has both longitudinal component $\boldsymbol{a}_{1l}$ and transverse component $\boldsymbol{a}_{1t}$. In the momentum space, $\boldsymbol{a}_{1l}$ and $\boldsymbol{a}_{1t}$ are defined as

\begin{equation}
    \boldsymbol{a}_{1l}(\boldsymbol{q})=\boldsymbol{q}\left(\boldsymbol{q}\cdot\boldsymbol{a}_1\right)/q,\label{eq:transverse}
\end{equation}
and
\begin{equation}
    \boldsymbol{a}_{1t}=\boldsymbol{a}_1-\boldsymbol{a}_{1l}.
\end{equation}
Due to the charge conservation, the total current density $\boldsymbol{j}_1$ has to be a transverse field with zero longitudinal component $\boldsymbol{\nabla}\times\boldsymbol{j}_1=0$. For convenience, we choose the Coulomb gauge for the $\boldsymbol{A}$ field, $\boldsymbol{\nabla}\cdot\boldsymbol{A}=0$. Thus the phase gradient term should cancel the longitudinal component of $\boldsymbol{a}$

\begin{equation}
    \boldsymbol{\nabla}\phi=2e\boldsymbol{a}_{1l}.
\end{equation}

Then the current density becomes

\begin{equation}
    \boldsymbol{j}=-e^2D\left(\boldsymbol{A}+\boldsymbol{a}_{1t}\right)\delta(z).\label{eq:current3}
\end{equation}

Combining the above expression of the current and the Maxwell equation we arrive at

\begin{equation}
\boldsymbol{\nabla}^2\boldsymbol{A}=-\mu_0\boldsymbol{j}=e^2D\mu_0\left(\boldsymbol{A}+\boldsymbol{a}_{1t}\right)\delta(z)
\end{equation}

In the momentum space, we have

\begin{equation}
    -q^2\boldsymbol{A}=e^2D\mu_0 \int dq_z\boldsymbol{A}+e^2D\mu_0 \boldsymbol{a}_{1t}. \label{eq:LondonMaxwell2}
\end{equation}
 
Moving $q^2$ to the right-hand side of Eq. (\ref{eq:LondonMaxwell2}) and integrating the equation over $q_z$ leads to 

\begin{equation}
    -K(q_{\parallel})=\frac{e^2\mu_0DK(q_{\parallel})}{2q_{\parallel}}+\frac{e^2\mu_0D\boldsymbol{a}_{1t}}{2q_{\parallel}}\label{eq:MaxWell3}
\end{equation}

with
 \begin{equation}
     K(q_{\parallel})=\boldsymbol{A}(z=0)=\frac{1}{2\pi}\int dq_z\boldsymbol{A},
 \end{equation}
and

\begin{equation}
    q_{\parallel}=\sqrt{q_x^2+q_y^2}.
\end{equation}

Solving Eq. (\ref{eq:MaxWell3}), we obtain

\begin{equation}
    K(q_{\parallel})=-\frac{\boldsymbol{a}_{1t}}{1+q_{\parallel}r_0},
\end{equation}
where $r_0$ is Pearl length \cite{pearl1964current} defined as

\begin{equation}
    r_0=\frac{2}{e^2\mu_0D}.
\end{equation}
Thus the current is given by

\begin{equation}
    \boldsymbol{j}_1=-e^2D\frac{q_{\parallel}r_0\boldsymbol{a}_{1t}}{1+q_{\parallel}r_0}.\label{eq:currentq}
\end{equation}
We perform the 2D Fourier transformation and obtain the current in the real space

\begin{equation}
    \boldsymbol{j}_1=\frac{e^2D}{2\pi}\left[\boldsymbol{a}_1\partial^2_{\boldsymbol{r}}-(\boldsymbol{a}_1\cdot\partial_{\boldsymbol{r}})\partial_{\boldsymbol{r}}\right]V(r),\label{eq:currentf}
\end{equation}
where $V(r)$ is the Keldysh potential givn by

\begin{equation}
V(r)=\frac{\pi}{2}\left[H_0\left(\frac{r}{r_0}\right)-Y_0\left(\frac{r}{r_0}\right)\right]\delta(z),
\end{equation}
where $H_0$ is the Struve function and $Y_0$ is the second kind Bessel function.  Substituting Eq. (\ref{eq:currentf}) into Eq. (\ref{eq:energyf}) we get the effective magnetic interaction between $\boldsymbol{m}_1$ and $\boldsymbol{m}_2$

\begin{equation}
   F_I=-\frac{e^2D}{2\pi}\left[\boldsymbol{a}_1\cdot\boldsymbol{a}_2\partial^2_{\boldsymbol{r}}-(\boldsymbol{a}_1\cdot\partial_{\boldsymbol{r}})(\boldsymbol{a}_2\cdot\partial_{\boldsymbol{r}})\right]V(r).
\end{equation}
In terms of $\boldsymbol{m}$, this interaction can be written as

\begin{equation}
    F_I=-Z\sum_{\alpha\beta}m_{1\alpha}m_{2\beta}\partial_{\alpha}\partial_{\beta}V(r),
\end{equation}
with $Z=e^2DJ_0^2\Gamma^2/2\pi$, $\alpha ,\beta\equiv x, y$. This is Eq. (6) in the main text. By the simplest possible matching of the two asymptotic behaviors we can construct an approximated expression for the Keldysh potential in terms of elementary functions \cite{cudazzo2011dielectric}

\begin{equation}
    V_{eff}=-\left[\log\left(\frac{r}{r+r_0}\right)+\left(r-\log2\right)e^{-r/r_0}\right].\label{eq:KeldyshpotentialS2}
\end{equation}
Using this approximate expression of the Keldysh potential,  we get the approximate interaction 
\begin{equation}
F_I=J_{\perp}\boldsymbol{m}_{1\perp}\boldsymbol{m}_{2\perp}-J_{\parallel}\boldsymbol{m}_{1\parallel}\boldsymbol{m}_{2\parallel}\;, 
\end{equation}
with

\begin{equation} 
    J_{\perp}=Z\frac{r_0}{r^2(r+r_0)}, \quad J_{\parallel}=Z\frac{r_0(2r+r_0)}{r^2(r+r_0)^2}.
\end{equation}
These are Eq. (8) and (9) in the main text. The supercurrent can be obtained by substituting Eq. (\ref{eq:KeldyshpotentialS2}) into Eq. (\ref{eq:currentf})
\begin{equation}
    \boldsymbol{j}(\boldsymbol{r}_j)=\sum_i\frac{1}{\Gamma  
 J_0}\left[J_{\parallel}\boldsymbol{m}_i\times\hat{z}-(J_{\parallel}+J_{\perp})(\boldsymbol{m}_i\times\hat{z}\cdot\hat{r}_{ij})\hat{r}_{ij}\right].
\end{equation}
This is Eq. (13) in the main text.

\section{S3. Derivation of superfluid weight in Rashba superconductors}
We consider a 2D Rashba superductor. We assume that the Rashba spin-orbit coupling is much larger than the pairing $\Delta$ so that one can project the Hamiltonian onto the two helical bands denoted by $\lambda=\pm$. This can be described by the following Eilenberger equation  \cite{houzet2015quasiclassical}

\begin{equation}
-v_F\boldsymbol{n}\cdot \left(\boldsymbol{\nabla}\hat{g}_{\lambda\boldsymbol{n}}-ie\boldsymbol{A}[\tau_3,\hat{g}_{\lambda \boldsymbol{n}} 
]\right)= [\omega_n\tau_3+\Delta \tau_1,\hat{g}_{\lambda \boldsymbol{n}}],
\end{equation}
where $\hat{g}_{\lambda\boldsymbol{n}}$ is the quasiclassical Green function of band $\lambda$ with momentum direction $\boldsymbol{n}$.  $\tau$ is the Pauli matrix acting on the particle-hole space. $\Delta$ is the pairing gap. $\omega_n$ is the Matsubara frequency $\omega_n=(2n+1)T$, with $n\in Z$ and $T$ denoting the temperature. $v_F$ is the Fermi velocity and $e$ is the electron charge. Soling the Eilenberger equation with the normalization condition $\hat{g}_{\lambda\boldsymbol{n}}^2=1$, we get the quasiclassical Green function

\begin{equation}
    \hat{g}_{\lambda\boldsymbol{n}}=\frac{(\omega_n-iv_Fe\boldsymbol{A}\cdot\boldsymbol{n})\tau_3+\Delta\tau_1}{\sqrt{\left(\omega_n-iv_Fe\boldsymbol{A}\cdot\boldsymbol{n}\right)^2+\Delta^2}}.\label{eq:Eilenberger1}
\end{equation}
For small gauge field, we can expand $\hat{g}_{\lambda\boldsymbol{n}}$ up to the first order in $\boldsymbol{A}$

\begin{equation}
    \hat{g}_{\lambda\boldsymbol{n}}\approx\frac{\omega_n\tau_3+\Delta\tau_1}{\sqrt{\omega_n^2+\Delta^2}}-\frac{iv_Fe\Delta^2\boldsymbol{A}\cdot\boldsymbol{n}\tau_3}{\sqrt{\omega_n^2+\Delta^2}^3}\label{eq:Eilenberger2}
\end{equation}

The supercurrent flowing in the superconductor is given by

\begin{equation}
    \boldsymbol{j}=-\frac{1}{2}i\pi e\sum_{\lambda,n}N_{\lambda}T\text{Tr}\langle v_F\boldsymbol{n}\tau_3\hat{g}_{\lambda\boldsymbol{n}}\rangle, \label{eq:currentEilenberger}
\end{equation}
where $\langle\cdot\rangle$ means the average over all the directions. $N_{\lambda}$ is the band-dependent density of states near the Fermi energy

\begin{equation}
    N_{\lambda}=\frac{N_0}{2}\left(1+\lambda\frac{\alpha}{v_F}\right),
\end{equation}
where $N_0$ is the total density of states. Substituting Eq. (\ref{eq:Eilenberger2}) into Eq. (\ref{eq:currentEilenberger}), we have

\begin{equation}
    \boldsymbol{j}=-\frac{1}{2}\pi e^2v_F^2N_0T\boldsymbol{A}\sum_n\frac{\Delta^2}{\sqrt{\omega_n^2+\Delta^2}^3}
\end{equation}
 The superfluid weight is given by

 \begin{equation}
     D=-\frac{1}{2}\frac{1}{e^2}\frac{\partial\boldsymbol{j}}{\partial \boldsymbol{A}}=2\pi v_F^2N_0T\sum_n\frac{\Delta^2}{\sqrt{\omega_n^2+\Delta^2}}.\label{eq:superfluidweight}
 \end{equation}
We use this superfluid weight to determine the Pearl length in Eq. in the main text.

\section{S4. Derivation of $\Gamma$ in Rashba superconductors at clean limit}
Here we consider a 2D Rashba superconductor with an exchange field described by the Eilenberger equation

\begin{equation}
-\boldsymbol{n}\cdot \left(v_F\boldsymbol{\nabla}\hat{g}_{\lambda\boldsymbol{n}}-ie\lambda\boldsymbol{h}\times\hat{z}[\tau_3,\hat{g}_{\lambda \boldsymbol{n}} 
]\right)= [\omega_n\tau_3+\Delta \tau_1,\hat{g}_{\lambda \boldsymbol{n}}].
\end{equation}
The solution of this Eilenber equation is

\begin{equation}
    \hat{g}_{\lambda\boldsymbol{n}}=\frac{(\omega_n-i\lambda e\boldsymbol{h}\times\hat{z}\cdot\boldsymbol{n})\tau_3+\Delta\tau_1}{\sqrt{\left(\omega_n-i\lambda e\boldsymbol{h}\times\hat{z}\cdot\boldsymbol{n}\right)^2+\Delta^2}}.
\end{equation}
Using the same method as we did in section S3, we get the anomalous current induced by the exchange field

\begin{equation}
    \boldsymbol{j}^a=-\frac{1}{2}\pi e^2\alpha\boldsymbol{h}\times\hat{z}N_0T\sum_n\frac{\Delta^2}{\sqrt{\omega_n^2+\Delta^2}}.\label{eq:anomalouscurrent}
\end{equation}
The $\Gamma$ parameter by definition is obtained from Eq. (\ref{eq:superfluidweight}) and Eq. (\ref{eq:anomalouscurrent})

\begin{equation}
    \Gamma=\frac{\alpha}{v_F^2}.
\end{equation}
This is what we show above Eq. (2) in the main text.

\end{document}